\documentclass{jpsj3}
\usepackage{txfonts}
\usepackage[dvipdfmx]{graphicx}
\title{Introduction of spin-orbit interaction into graphene with hydrogenation}

\author{Taketomo Nakamura$^1$\thanks{taketomo@issp.u-tokyo.ac.jp}, 
Junji Haruyama$^2$,\\ and Shingo Katsumoto$^1$}
\inst{$^1$Institute for Solid State Physics, University of Tokyo, 5-1-5 Kashiwanoha, Kashiwa, Chiba 277-8581, Japan\\
$^2$Department of Electrical Engineering and Electronics, Aoyama Gakuin University,
5-10-1 Fuchinobe, Sagamihara, Kanagawa 252-5258, Japan} 

\abst{Introduction of spin-orbit interaction (SOI) into graphene with weak hydrogenation ($\sim$0.1\%) by dissociation of hydrogen silsesquioxane resist has been confirmed through the appearance of inverse spin Hall effect. The spin current was produced by spin injection from permalloy electrodes excluding non-spin relating experimental artifact.}

\newcommand{\figwidth}{0.84\linewidth}
\begin{document}
\maketitle
A virtual graphene with strong spin-orbit interaction (SOI) is first considered as a model system of topological insulator\cite{PhysRevLett.95.226801} though the SOI in actual graphene is almost zero reflecting the atomically planer structure\cite{Konschuh}.
Recently there has been a report on huge enhancement of SOI in lightly hydrogenated graphene\cite{balakrishnan2013colossal}, in which work, the introduction of hydrogen was done by dissociation of a hydrogen silsesquioxane (HSQ). Because their experiment is restricted to non-local transport measurement, more spin-specific confirmation is desirable. In this short note, we report observation of inverse spin Hall effect in lightly hydrogenated graphene caused by spin current injected from a permalloy (Py) electrode.
The results support the claim of strong enhancement of SOI by hydrogenation in Ref.\citen{balakrishnan2013colossal}.

\begin{figure}[b]
\centering
\vspace{-4mm}
\includegraphics[width=\figwidth,clip]{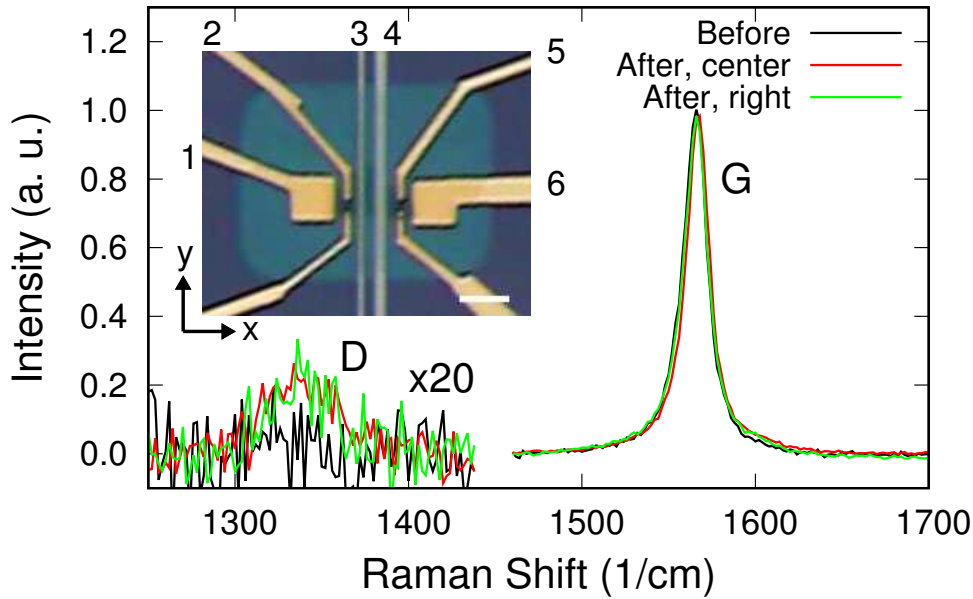}\hfil
\caption{Raman spectra of the graphene before and after hydrogenation. The signals for D mode are magnified by 20.
The inset displays optical micrograph of the sample with the terminal numbers. The scale bar represents 5 $\mu$m. 
}
\label{fig_sample}
\end{figure}
We used a graphene exfoliated on a SiO$_2$ substrate from a highly oriented pyrolytic graphite.
Two strips of Py (Ni 81\%, Fe 19\%) with the widths of 0.5~$\mu$m and 2.0~$\mu$m, the thickness of 80~nm were formed with e-beam lithography and vacuum deposition, being indicated as electrodes 3 and 4 in the inset of Fig.\ref{fig_sample}. Non-magnetic Ti/Au electrodes numbered 1, 2, 5, 6 with shapes displayed in the inset of Fig.\ref{fig_sample} were also fabricated with e-beam lithography.
HSQ resist was spin-coated and 75~keV electron beam was scanned over the sample area for the hydrogenation through dissociation.
The dose density was 10~mC/cm$^2$.

Figure \ref{fig_sample} shows Raman spectra of the graphene before and after the hydrogenation. Apparent relative enhancement in D mode is observed, and from the ratio D/G the amount of hydrogenation is estimated to be 0.1\%\cite{balakrishnan2013colossal}. The Raman shift of G mode indicates $n$-type doping\cite{Ferrari2007}.

\begin{figure}
\centering
\includegraphics[width=\figwidth,clip]{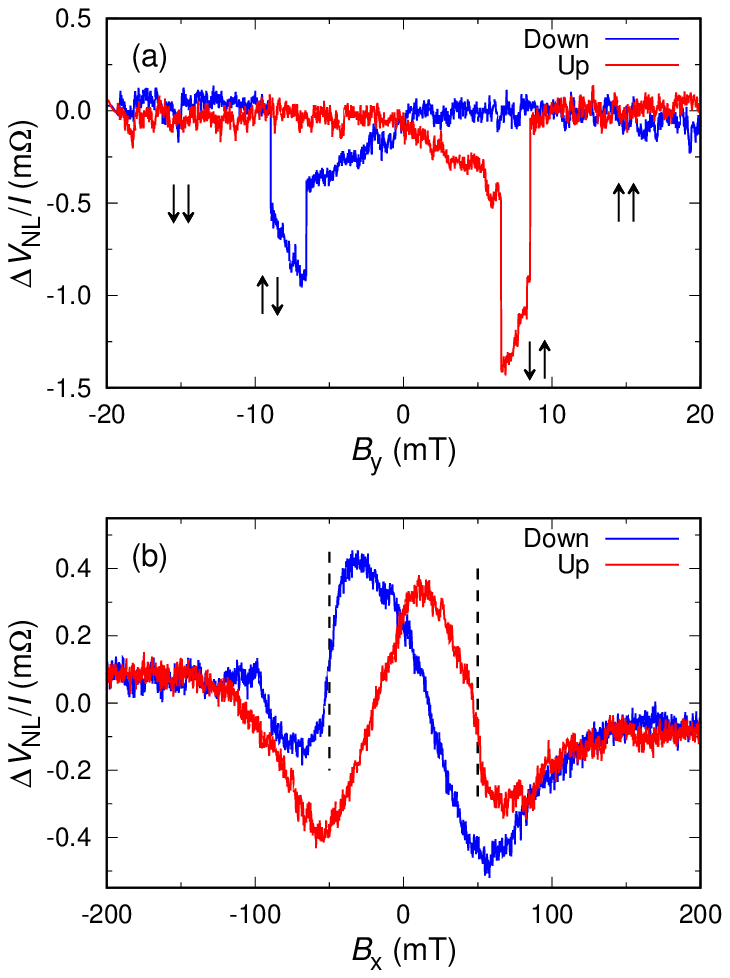}
\caption{(a) Spin valve magnetoresistance for the current flow between terminals 4 and 6, the voltage between terminals 1 and 3. The magnetic field is applied along the Py strips ($y$-direction). The temperature is 43~K. The pairs of arrows indicate the magnetization directions of the Py strips.
(b) ``Hanle-type'' magnetoresistance observed at 7~K. The terminal configuration is the same as in (a). The in-plane external magnetic field direction is perpendicular to the Py strips ($x$-direction). Vertical broken lines indicate regions of rapid change in $\Delta V_\mathrm{NL}$
}
\label{fig_valve}
\end{figure}

Figure \ref{fig_valve}(a) shows non-local spin-valve magnetoresistance measured at 43~K in the configuration of current (3.5~mA) between terminals 4 and 6, and voltage between terminals 1 and 3. The external magnetic field direction was along the Py strips.
From the measurement of anisotropic magnetoresistance in Py strips, we confirmed that the coercive forces were 6$\sim$7~mT and 8$\sim$9~mT for wider (4) and narrower (3) strips respectively, in accordance with the jump fields of the spin-valve magnetoresistance. Though the line-shape contains distortions probably due to spins with weak coercivity, {\it i.e.} weak pinning force and residual unflipped magnetic domains in Py, this observation certifies that spin-polarized electrons were injected into the graphene layer, since the observed spin-valve signal of $\varDelta R_{\rm s}$ = 0.98~m$\Omega$ is proportional to the injected spin current and the inner product of magnetizations in the two ferromagnetic electrodes $\boldsymbol{M}_\mathrm{i}\cdot\boldsymbol{M}_\mathrm{d}$.
The spin injection efficiency $P$ is estimated to be 0.1\% for this device, assuming $\sigma_\square = 0.5$~mS\cite{Jozsa2009}.

In the same terminal configuration, we measured the magnetoresistance for the magnetic field directed along $x$-axis indicated in the inset of Fig.\ref{fig_sample}, {\it i.e.}, in the plane of graphene sheet and perpendicular to the Py strips.
The temperature here and henceforth was 7~K.
Unfortunately, as mentioned above, the pinning force of the in-plane magnetic field is so weak in the present Py and the magnetization seems to be significantly slanted to $y$-axis by the magnetic field though the strips were magnetized along $y$ beforehand.
This appeared as significant hysteresis in the magnetoresistance.

\begin{figure}
\centering
\includegraphics[width=\figwidth,clip]{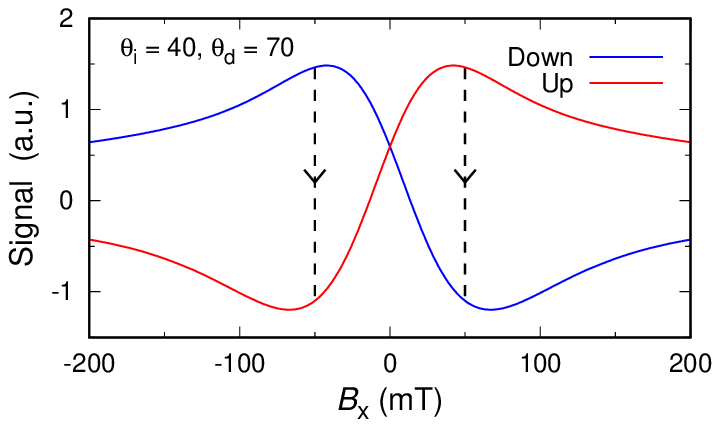}
\caption{Non-local Hanle magnetoresistance calculated on Eq.\eqref{eq_diffusion_hanle} with $B_\mathrm{eff}= 60$~mT, $D=10^{-3}$~m$^2$/s, $L=1$~$\mu$m, $\tau=100$~ps, and the other parameters written in the figure.
Vertical broken lines indicate positions of rapid variation in the observed (Fig.\ref{fig_valve}) line-shapes.}
\label{fig_hanlecal1}
\end{figure}
In spite of these unfavorable distortions, we still see distinct Hanle features in the signal. 
Attention should be paid to the fact that the maxima are shifted from zero-field. 
Because the external magnetic field in the present configuration causes out-of-plane spin rotation, 
maxima of non-local resistance should appear at zero-field for electrons {\it without} SOI.
In a diffusive transport model, the non-local resistance $\varDelta R$ due to spin current is written as
\begin{equation}
\varDelta R=\frac{P^2}{e^2nA}\int_0^\infty\!
\frac{dt}{\sqrt{4\pi Dt}}\exp\left(-\frac{L^2}{4Dt}-\frac{t}{\tau}\right)f(\boldsymbol{M}_\mathrm{i},\boldsymbol{M}_\mathrm{d},B,t),
\label{eq_diffusion_hanle}
\end{equation}
where $A$ is the junction area, $D$ the diffusion constant, $L$ the detector-injector distance, $\tau$ the spin relaxation time.
The functional form of $f$ reflects the configuration of magnetization and the SOI.
Let $\theta_\mathrm{i,d}$ be the angle of $\boldsymbol{M}_\mathrm{i,d}$ measured from initial direction of external field $B$, $\omega_{\rm L}$ the Larmor frequency, 
$\theta_B=\tan^{-1}B_{\rm eff}/B$, where $B_{\rm eff}$ is the effective magnetic field originating from Rashba-type SOI.
Then $f$ is proportional to 
\[
\sin(\theta_B-\theta_\mathrm{i})\sin(\theta_B-\theta_\mathrm{d})\cos\omega_\mathrm{L}t
+\cos(\theta_B-\theta_\mathrm{i})\cos(\theta_B-\theta_\mathrm{d}).
\]

In Fig.\ref{fig_hanlecal1} we show an example of line-shape calculated on the model Eq.\eqref{eq_diffusion_hanle}, where we adopt $B_{\rm eff}$=60~mT.
In this model, the magnetizations of Py strips are fixed for simplification, which apparently did not hold in the experiment.
If one assumes rapid changes in magnetization directions at the positions indicated by vertical broken lines, the combined line-shapes well resemble the observed ones in Fig.\ref{fig_valve}(b) supporting the existence of Rashba-type SOI in the graphene\cite{balakrishnan2013colossal}.
The above observations in Fig.\ref{fig_valve} thus clearly confirm that the injected spins transported over 1~$\mu$m
with remarkable coherence despite the emergence of a strong SOI.

\begin{figure}
\centering
\includegraphics[width=\figwidth,clip]{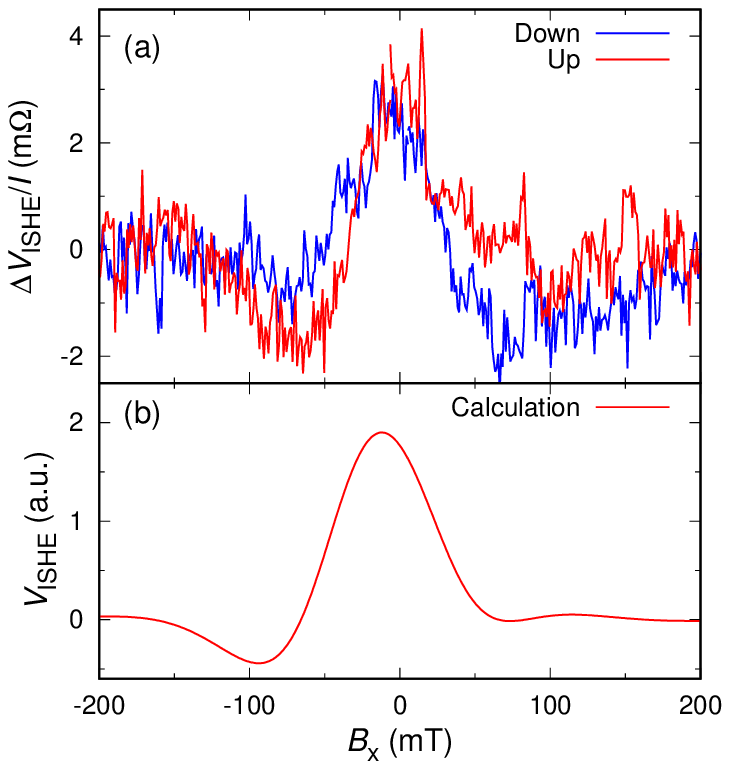}
\caption{(a) Inverse spin Hall signal measured at electrodes 5 and 6 as a function of magnetic field. 
The current between 1 and 4 terminals was 4~mA.
(b) Calculated line-shape of inverse spin Hall voltage $V_\mathrm{ISHE}$ with the same parameters as Fig.\ref{fig_hanlecal1}, 
assuming $V_\mathrm{ISHE}$ proportional to $S_z$ at the probes.
}
\label{fig_invspinhall}
\end{figure}

Lastly we changed the current terminals to 1 and 4 and measured the voltage between 5 and 6.
Though there was neither net current nor magnetic material in the probe region, clear magnetization-originated signal appeared as shown in Fig.\ref{fig_invspinhall}(a).
Since the SOI inside the Au terminals cannot flip up the spins,
the above behavior can only be explained by inverse spin Hall effect\cite{:/content/aip/journal/apl/88/18/10.1063/1.2199473}, which arose from the Rashba type SOI in the graphene layer and the pure spin current from the Py strip.
The signal in Fig.\ref{fig_invspinhall}(a) takes maxima at around $B=0$ unlike that in Fig.\ref{fig_valve}(b).

Let $\boldsymbol{J}_{\rm s}$ and $\boldsymbol{S}$ be the flux and the spin of the spin current respectively and the inverse spin Hall current $\boldsymbol{J}_{\rm c}$ is proportional to $\boldsymbol{J}_{\rm s}\times\boldsymbol{S}$. 
Since $\boldsymbol{J}_{\rm s}$ was along $x$-axis and the terminal 5 probed the voltage along $y$-axis, we can express $V_\mathrm{ISHE}$ proportional to $S_z\propto \sin(\theta_B-\theta_\mathrm{i})\sin\omega_{\rm L}t$.
Then the same parameters as those in Fig.\ref{fig_hanlecal1} reproduce the lineshape with the maximum at around $B=0$ as shown in Fig.\ref{fig_invspinhall}(b), supporting legitimacy of the present analysis. 
The lower bound in the SOI strength is estimated from the spin Hall angle $\Delta R_\mathrm{ISHE}\sigma_0/P \sim 0.02$, roughly corresponding to $\Delta_\mathrm{SOI}\sim 0.1$~meV, 10 times larger than the value of pristine graphene\cite{Konschuh}. This value is still smaller than that reported in Ref.\citen{balakrishnan2013colossal} though may jump up with more accurate estimation of {\it e.g.}, the form factor etc.

In summary, we have confirmed the introduction of Rashba-type SOI into graphene with e-beam dissociation of HSQ resist.

\begin{acknowledgment}
This work was supported by Grant-in-Aid for Scientific Research on Innovative Area, ``Nano Spin Conversion Science" (Grant No.26103003),
also by Grants No.25247051 and No.15K17676,
by Special Coordination Funds for Promoting Science and Technology,
and by Grant for Basic Science Research Projects from The Sumitomo Foundation.
\end{acknowledgment}

\end{document}